\documentclass[a4paper,10pt]{article}

\usepackage{contribution}

\newcommand{\weblink}[2][]{%
    \ifthenelse{\equal{#1}{}}%
    {\textnormal{\url{#2}}}%
    {\textnormal{\href{#2}{#1}}}%
}

\def\beq{\begin{equation}}
\def\eeq#1{\label{#1}\end{equation}}
\def\eeqn{\end{equation}}

\def\beqa{\begin{eqnarray}}
\def\eeqa#1{\label{#1}\end{eqnarray}}
\def\eeqan{\end{eqnarray}}

\let\bar=\overbar

\def\D{{\cal D}}

\def\Dslash{\not{\hbox{\kern-4pt $D$}}}
\def\dslash{\not{\hbox{\kern-2pt $\del$}}}

\def\msb{{\bar{\ssstyle M \kern -1pt S}}}

\newcommand{\contribution}[7][]{%
  \clearpage
  \thispagestyle{plain}

  \ifthenelse{\equal{#1}{}}
  {\hypersetup{pdftitle={#2}}}
  {\hypersetup{pdftitle={#1}}}
  \hypersetup{pdfauthor={{#3} {#4}}}
  {\centering\normalfont\LARGE\bfseries\sffamily #2 \par\nobreak}
  \lhead{}
  \chead{%
    \textit{\footnotesize XXIInd International Workshop ``High-Energy Physics and Quantum Field Theory'', 
June 24 -- July 1, 2015, Samara, Russia}%
  }
  \rhead{}
  \bigskip
  \begin{center}
    {#3} {#4}\ifthenelse{\equal{#6}{}}{}{\footnote{\weblink[#6]{mailto:#6}}}
    \ifthenelse{\equal{#7}{}}{}{#7} \\
    \textit{#5}
  \end{center}
  \bigskip
}

\renewcommand{\abstract}[1]{%
  \begin{center}
    \begin{minipage}{0.85\textwidth}
      \begin{footnotesize}
        #1
      \end{footnotesize}
    \end{minipage}
  \end{center}
  \bigskip
}

\begin{document}

%
%
%
%
%
%
{  


%

\contribution[Radiative decays $V\rightarrow P\gamma^{*}$ in the instant form of relativistic quantum mechanics]  
{Radiative decays $V\rightarrow P\gamma^{*}$ in the instant form of relativistic quantum mechanics}  
{Alexander}{Krutov,}  
{Samara State University, \\
Academician Pavlov St.,1, 443011 Samara, Russia\\D.V.~Skobeltsyn
Institute of Nuclear Physics,
M. V. Lomonosov Moscow State University, 119991 Moscow, Russia} 
{krutov@samsu.ru}  
{Roman Polezhaev, Vadim Troitsky}  
%

\abstract{%
Calculations of form factor for the radiative decays $V\rightarrow
P\gamma^{*}$ process are performed in the framework of an instant
form of relativistic quantum mechanics. The electromagnetic
current operator for this decay is constructed. The transition
form factor is obtained in the so called relativistic modified impulse
approximation (MIA). The current operator satisfies the conditions
of Lorentz-covariance and current conservation in MIA. The results of
the calculations are compared with the analogous results in the
light-front dynamics and in the model of vector meson dominance}
%

\section{Introduction}
Understanding the relation between the observable properties of
mesons or baryons and the underlying dynamics is a serious
challenge in modern hadron physics. Quantum chromodynamics is a
good candidate for the role of a complete theory. Moreover, it is
known that the perturbative calculation methods in QCD yield
reliable results only when describing so-called "hard" processes
characterized by large transferred momenta and fail when
calculating characteristics determined by "soft" processes. This,
for instance, pertains to describing particle electromagnetic form
factors and other electromagnetic characteristics that are measured
experimentally. The field theory of strong interactions involves
infinitely many degrees of freedom transferred by local fields -
quarks and gluons. But the success of constituent models indicates
that the main characteristics of hadronic states can be described
using only a finite number of these degrees of freedom, leaving
other degrees of freedom frozen. One of the most reliable methods
for such a description is to use the relativistic quantum
mechanics (RQM) model. As in every other theoretical description
of electromagnetic structures of composite systems, we have three
main points in the relativistic constiruent model.

1. First, we must choose the dynamics. Foundations of RQM were
laid by Dirac in \cite{Dir49}, where three main types of dynamics
differing by evolution parameters were introduced. Those are the
point form of dynamics, the instant form of dynamics, and the
light-front dynamics, each of which can be related to a
three-dimensional hypersurface in the four-dimensional space. The
initial data are defined on this hypersurface, and its evolution
is the main object of study. A point form hypersurface is
determined by the conditions $- x^{\mu}x_{\mu} = a^{2}$, $t > 0$,
an instant form hypersurface is determined by the condition $t =
0$, and a light-front form hypersurface is determined by the
equation $x^{0} + x^{3} = 0$.

Numerous actual calculations based on the Dirac model currently
exist. These calculations illustrate the power and flexibility of
the RQM as an instrument for studying systems of strongly
interacting particles in the energy range below several GeV (see
\cite{LeS78,KeP91,Coe92,GiG02,KrT09,PoE11,Pol89} for a discussion
and relevant references).

2. Second, we must construct the operator of the transition
current. This step is important in all schemes for studying
particle structures, not only in RQM. Separating the kinematic
parts and invariant parts (form factors) of the matrix elements of
the current is an important point in all such methods. We call a
representation of a current matrix element in terms of form
factors a parameterization of the matrix element. We note that
form factors, which are Lorentz-invariant functions, are
customarily determined from experimental data. Therefore, we must
consider a matrix element of the electroweak current that has
correct transformation properties, and we must perform the
procedure for relativistic parameterization of the current matrix
element, i.e., we must separate relativistic-invariant form
factors or, in other words, separate reduced matrix elements using
the Wigner-Eckart theorem for the Poincare group \cite{KrT05}. An
important feature of composite systems is that all these form
factors, strictly speaking, are distributions or generalized
functions.

3. Third, choosing a concrete approximation for calculations is
important. As a rule, we use the impulse approximation in the case
of two-component systems, which means that a virtual
$\gamma$-quantum interacts with only one of the components, not
with both of them simultaneously. But when choosing a
approximation, we must always verify that it does not break the
relativistic invariance. In the instant form of RQM, the impulse
approximation is not relativistic invariant and depends on the
choice of the coordinate system. To ensure the relativistic
invariance, we construct a modified impulse approximation (MIA) in
the spirit of \cite{TrS69}. This method uses a double-integral
representation with respect to the invariant mass of the system.
These integrals appear as a result of solving equations of the
Muskhelishvili-Omnes type. This method was laid out in detail in
\cite{KrT02} (see Secs. 3B and 3C), and its use in calculating
structures of concrete two-quark systems can be found in
\cite{BaK95,BaK96,KrT99}. In particular, the method was used to
calculate the pion form factor \cite{KrT01}, both in the
asymptotic domain \cite{KrT98} and in the timelike domain
\cite{KrN13}. We emphasize that the calculations in \cite{KrT01}
also demonstrated their predictive power (see \cite{KrTT09,TrT13}
for the details).

In this paper we are consider the description of radiative decays
$V\rightarrow P\gamma^{*}$ in instant form of (RQM). Construction
of electromagnetic current operator that satisfies the
Lorentz-covariance and current conservation conditions is
performed. We calculate the transition form factor and compare it
with results of other approaches.

\section{Parameterization an electromagnetic current matrix element nondiagonal in the angular
momentum in the case of a free two-particle system with quantum numbers of pions and
$\rho$-mesons.}

The state vector of the system of two interacting particles in the
RQM belongs to the direct product of two single-particle Hilbert
spaces. Therefore, as a bases, we can consider the following two
sets of vectors.

1. The basis of the individual momenta and spins of the particles:
\begin{eqnarray}
\label{dv}
&& \hspace{-2mm}|\vec{p_1},m_1;\vec{p_2},m_2\rangle = |\vec{p_1},m_1\rangle
\otimes{|\vec{p_2},m_2}\rangle\;,\nonumber\\[2mm]&&\langle\vec p, m|\vec
p\;',m'\rangle = 2\,p_0\,\delta(\vec p - \vec
p\;')\,\delta_{m\,m'}\;,
\end{eqnarray}
where $\vec p_{1,2}$ - three-momentum, $m_{1,2}$ -spin projection, $p_0^2 - \vec p\,^2 = M^2$.

2. Basis with separated center of mass motion of two particles:
\begin{equation}
\label{vec}
|\vec{P},\sqrt{s},J,L,S,m\rangle\;,
\end{equation}
where $\vec{P} = \vec{p_{1}}+{\vec{p_{2}}}$,  $\sqrt{s}$ -
invariant mass of two particle system, $P^2 = s,\;$
$L$- orbital momentum, $S$-spin momentum.

Basis (\ref{dv}) and (\ref{vec}) are related by the Clebsh-Gordon
decomposition:
\begin{eqnarray}\label{Clebsh111}
&& \hspace{-2mm} |\vec{P},\sqrt{s},J,L,S,m\rangle =
\sum_{m_{1},m_{2}}\int\,\frac{d^3{\vec{p_1}}}{2p_{10}}\frac{d^3{\vec{p_2}}}{2p_{20}}
|\vec{p_1},m_{1};\vec{p_2},m_{2}\rangle
\nonumber\\[1mm]&& \times\langle\vec{p_1},m_{1};\vec{p_2},m_{2}|\vec{P},\sqrt{s},J,L,S,m\rangle\;,
\end{eqnarray}
where
\begin{eqnarray}\label{cli}
&&\hspace{-2mm}
\langle\vec{p_1},m_{1};\vec{p_2},m_{2}|\vec{P},\sqrt{s},J,L,S,m\rangle
= \frac{2\sqrt{s}}{\sqrt{\lambda(s,M^2,M^2)}}
\nonumber\\
[2mm]&& \hspace{-2mm} \times 2P_{0}\delta(P-p_{1}-p_{2}) \,
\sum_{\tilde{m_{1}},\tilde{m_{2}}}D^{1/2}_{m_1\tilde{m}_1}(p_{1},P)
D^{1/2}_{m_2\tilde{m}_2}(p_{2},P)\, \nonumber\\
[2mm]&& \hspace{-2mm}\times\langle
\frac{1}{2}\tilde{m_{1}}\frac{1}{2} \tilde{m_{2}}|S m_{S}\rangle
Y_{L m_{L}}(\vartheta,\varphi)\langle S L m_{S} m_{L}|J
m\rangle\;,
\end{eqnarray}
$\lambda(a,b,c) = a^2+b^2+c^2-2(ab+ac+bc)\;$,$D^{1/2}$ - matrix of
three-dimensional rotations (Wigner's $D$-function),$\;Y_{L
m_{L}}$ -- are spherical functions.

In the constituent quark model pion and $\rho$-meson are
represented by bound states of $u$ and $\bar{d}$-quarks with the
quantum numbers $J=L=S=0$ for the pion and $J=S=1;L=0$ for the
$\rho$-meson. We omit zero-values quantum numbers in the
corresponding state vectors. Equating the quark masses,
$M_{u}=M_{\bar{d}}=M$, we can write the matrix element of the
electromagnetic current operator for the free two-particle system
with the given quantum numbers in the following form:

\begin{eqnarray}\label{Clebsh}
&& \hspace{-2mm}
\langle\vec{P},\sqrt{s}|j^{0}_\mu(0)|\vec{P}\;',\sqrt{s'},1,0,1,m'\rangle
=\int\,\frac{d^3{\vec{p_1}}}{2p_{10}} \frac{d^3{\vec{p_2}}}{2p_{20}}
\frac{d^3{\vec{p_1}'}}{2p'_{10}}
\frac{d^3{\vec{p_2}'}}{2p'_{20}}\cdot\nonumber\\[2mm]&& \hspace{-2mm}
\cdot
\langle\vec{P},\sqrt{s}|\vec{p_1},m_{1};\vec{p_2},m_{2}\rangle\,
\langle\vec{p_1},m_{1};\vec{p_2},m_{2}|j^{0}_\mu(0)|\vec{p_1}',m'_{1};\vec{p_2}',m'_{2}\rangle
\cdot\nonumber\\[2mm]&& \cdot\langle\vec{p_1}',m'_{1};\vec{p_2}',m'_{2}|\vec{P}\;',\sqrt{s'},1,0,1,m'\rangle\;,
\end{eqnarray}
where
\begin{eqnarray}\label{n1}
&& \hspace{-1mm}
\langle\vec{p_1},m_{1};\vec{p_2},m_{2}|j^{0}_\mu(0)|\vec{p_1}',m'_{1};\vec{p_2}',m'_{2}\rangle
= \langle\vec{p_1},m_{1}|j^{0}_{\mu 1}(0)|\vec{p_1}',m'_{1}\rangle
\cdot \nonumber\\[1mm]
&& \hspace{-1mm} \cdot \delta(\vec{p_2}-\vec{p_2}')\,\delta_{m_{2}
m_{2'}} + \langle\vec{p_2},m_{2}|j^{0}_{\mu
2}(0)|\vec{p_2}',m'_{2}\rangle \delta(\vec{p_1}-\vec{p_1}')\,
\delta_{m_{1} m_{1'}}\;.
\end{eqnarray}

Using nondiagonal parameterization of the matrix element
\cite{ChS63, KrP15} with the zero value of the total angular
momentum of the pion taken into account, we can write the matrix
element of the current by following way:

\begin{eqnarray}\label{A11}
&&\hspace{-2mm}
\langle\vec{P},\sqrt{s}|{j_0}(0)|\vec{P}\;',\sqrt{s'},1,0,1,m'\rangle
= \nonumber\\
[2mm]&& \hspace{-2mm}=\sum_{ \tilde{m}', l', k'} D^{1}_{m',\tilde{m}'}(P',w)
\langle 1 \tilde{m}' l'k' |0 0\rangle
\,Y_{l' k'}(\vec{q})\, G^{0,l'}_{01}(s,Q^2,s')\;,
\end{eqnarray}
\begin{eqnarray}\label{A111}
&&\hspace{-2mm}
\langle\vec{P},\sqrt{s}|{j}^{1}_t(0)|\vec{P}\;',\sqrt{s'},1,0,1,m'\rangle
= \nonumber\\
[2mm]&& \hspace{-2mm}=\sum_{\tilde{m}',l,k,j,n}
D^{1}_{m',\tilde{m}'}(P',w) \langle 1\tilde{m}' j n |0
0\rangle\langle 1 t l k |j n\rangle \,Y_{l
k}(\vec{q})\,G^{1,l,j}_{01}(s,Q^2,s')\;.
\end{eqnarray}

Comparing expressions (\ref{Clebsh}) and (\ref{A11}), (\ref{A111})
and integrating in the Breit reference system with
$\vec{q}=(0,0,q)$, we obtain analytic expressions for the so
called free two-particle form factors. These expressions are very
cumbersome, and we therefore write only one form factor, which we
need in what follows, explicitly:
\begin{eqnarray}\label{sv}
&&\hspace{-6mm}G_{01}^{111}(s,Q^2,s') =
\frac{\,\Theta(s,Q^2,s')(s+s'+Q^2)^{2}}{\sqrt{2}\sqrt{s-4M^2}
\sqrt{s'-4M^2}\sqrt{4M^2+Q^2}[\lambda(s,-Q^2,s')]^{1/2}}\cdot
\nonumber\\[2mm]&&\hspace{-6mm}
\cdot\,{\cos({\omega_1}+{\omega_2})/2}
\left(\frac{s'(s'-s+3Q^2)}{[\lambda(s,-Q^2,s')]^{1/2}}
(G_{M}^{u}(Q^2)+G_{M}^{\bar{d}})(Q^2)\right.)+\nonumber\\[2mm]&&\hspace{-6mm}
+ {\sin({\omega_1}+{\omega_2})/2}
\left(\frac{(s'-s-Q^2)}{(s+s'+Q^2)} \frac{\xi(s,s',Q^2)}{\sqrt{s'}}
(G_{M}^{u}(Q^2)+G_{M}^{\bar{d}}(Q^2))\right.-\nonumber\\[2mm]&&\hspace{-6mm}
-\left(\xi(s,s',Q^2)\frac{4M}{(s+s'+Q^2)}(G_{E}^{u}(Q^2)+G_{E}^{\bar{d}}(Q^2))
\right.)\;,
\end{eqnarray}

where: $\Theta(s,Q^2,s')=\vartheta(s'-s_{1})-\vartheta(s'-s_{2}),$
$\xi(s,s',Q^2)=\sqrt{ss'Q^2-M^2\lambda(s,-Q^2,s')},$
$$s_{1,2}=2M^{2}+\frac{1}{2M^2}(2M^2+Q^2)(s-2M^2)\mp\frac{1}{2M^2}\sqrt{Q^{2}(Q^{2}+4M^{2})s(s-4M^2)}\;,$$
$\vartheta$ - is the step function;

$$\omega_1=\arctan\frac{\xi(s,s',Q^2)}{M[(\sqrt{s}+\sqrt{s'})^2+Q^2]+\sqrt{ss'}(\sqrt{s}+\sqrt{s'})}\;,$$
$$\omega_2=\arctan\frac{\xi(s,s',Q^2)(2M+\sqrt{s}+\sqrt{s'})}{M(s+s'+Q^2)(2M+\sqrt{s}+\sqrt{s'})+\sqrt{ss'}(4M^2+Q^2)}\;.$$

\section{The transition form factor $ F_{ \pi \rho}(Q^2)$}

We can write the matrix element of the electromagnetic current of
the transition $\rho\rightarrow\pi\gamma^*$ as (see, e.g.,
\cite{CaG95}):

\begin{equation}\label{tok}
\langle{\vec {P}_{\pi}}|j^c_\mu(0)|{\vec { P}_{\rho}},1,{ m_{\rho}}\rangle
= F_{\pi\rho}(Q^2) \varepsilon_{\mu \nu \sigma \delta}
\xi^{\nu}(m_{\rho}){ P}_{\pi}^{\sigma}{ P}_{\rho}^{\delta}\;,
\end{equation}
where $\vec P_{\pi}$ and $\vec P_{\rho}$ are the three-momentum of
the pion and $\rho$-meson respectively, $\xi^{\nu}(m_{\rho})$ is
the polarization four-vector, $\varepsilon_{\mu \nu \sigma
\delta}$ is the rank-four anti-symmetric tensor, and
$F_{\pi\rho}(Q^2)$ is the transition form factor measured
experimentally.

For subsequent treatment of matrix element (\ref{tok}), we pass to
the Breit reference system. In this system, the polarization
vector is

\begin{equation}\label{pol}
\xi^{\nu}(m_{\rho})=-\frac{1}{\sqrt{2}}(0,0,1,i)\;.
\end{equation}

Substituting expression (\ref{pol}) in (\ref{tok}), we obtain

\begin{equation}\label{tok1}
\langle{\vec{\tilde P}}_{\pi}|{\tilde j^c}_1(0)|\vec{{\tilde
P}}_{\rho},1,\tilde{m_{\rho}}\rangle = -\frac{q ({\tilde P}_{\pi}^{0} +
{\tilde P}_{\rho}^{0})}{\sqrt{2}}F_{\pi\rho}(Q^2)\;.
\end{equation}

We note that the components $\tilde j^{c}_{0}, \tilde 
j^{c}_{2}$ and $\tilde j^{c}_{3}$ of current matrix element
(\ref{tok}) vanish. On the other hand, we can write matrix element
(\ref{tok}) using the procedure presented in the Section 2 for
parametrization of the current matrix element that is nondiagonal in
the total angular momentum:

\begin{eqnarray}\label{A11A}
&&\hspace{-2mm}
\langle\vec{P_{\pi}}|{j^{c}_0}(0)|\vec{P_{\rho}},1,m_{\rho}\rangle
= \nonumber\\
[2mm]&& \hspace{-2mm}=\sum_{ \tilde{m_{\rho}}, l', k'}
D^{1}_{m_{\rho},\tilde{m_{\rho}}}(P_{\rho},w)
 \langle 1 \tilde{m_{\rho}} l'k' |0
0\rangle \,Y_{l' k'}(\vec{q})\,G^{0,l'}_{01}(Q^2)\;,
\end{eqnarray}
\begin{eqnarray}\label{A111A}
&&\hspace{-2mm}
\langle\vec{P_{\pi}}|{j}^{c\,1}_t(0)|\vec{P_{\rho}},1,m_{\rho}\rangle
= \nonumber\\
[2mm]&& \hspace{-2mm}=\sum_{\tilde{m_{\rho}},l,k,j,n}
D^{1}_{m_{\rho},\tilde{m_{\rho}}}(P_{\rho},w)
  \langle 1\tilde{m_{\rho}} j n |0
0\rangle \,\langle 1 t l k |j n\rangle \,Y_{l
k}(\vec{q})\, G^{1,l,j}_{01}(Q^2)\;.
\end{eqnarray}

As mentioned above, for the radiative transition
$\rho\rightarrow\pi\gamma^*$, we have a unique form factor
expressed in terms of the first component of the electromagnetic
current matrix element:

\begin{equation}\label{tok2}
\langle{\vec{\tilde P}_{\pi}}|j^{c}_1(0)|{\vec{\tilde P}_{\rho}},1,{\tilde m}_{\rho}\rangle
= -\frac{1 }{\sqrt{3}}G_{01}^{111}(Q^2),
\end{equation}

where $G_{01}^{111}(Q^2)$ is the transition form factor in RQM.

Equating expressions (\ref{tok1}) and (\ref{tok2}), we obtain the
relation between experimentally measured form factor (\ref{tok})
and form factor (\ref{tok2}) arising in our parametrization
procedure:

\begin{equation}\label{transition}
F_{\pi\rho}(Q^2) = \sqrt{\frac{2}{3}} \frac{1}{q({\tilde P}_{\pi}^{0} +
{\tilde P}_{\rho}^{0})}G_{01}^{111}(Q^2)\;,
\end{equation}
where ${\tilde P}_{\pi}^{0}=\sqrt{M_{\pi}^{2}+{\vec
q}\;^2}\;,\quad {\tilde P}_{\rho}^{0}=\sqrt{M_{\rho}^{2} + {\vec
q}\;^2}.$ $\;M_\pi ,\; M_\rho$ are the masses of the pion and
$\rho$-meson respectively. For further calculations, we need an
analytic expression for the composite system form factor
$G_{01}^{111}(Q^2)$ in terms of wave functions in RQM.

\section{The expression for the transition form factor in the instant
form of RQM}

Let us again consider matrix element (\ref{tok}) of the operator of the transition electromagnetic current. We
assume that we have two-quark systems with quantum numbers corresponding to pions and $\rho$-mesons.
Because a state vector of a two-particle system in RQM belongs to the direct product of two one-particle
Hilbert spaces, we can expand matrix element (\ref{tok}) us:
\begin{eqnarray}\label{nor1}
&&\hspace{-2mm} \langle
\vec{P_{\pi}}|j_{\mu}^{c}(0)|\vec{P_{\rho}},1,m_{\rho}\rangle
=\sum \int\,\frac{d^3\vec{P}}{N} \frac{d^3\vec{P}'}{N'} d\sqrt{s}
d\sqrt{s'} \langle\vec{P_{\pi}}|\vec{P},\sqrt{s}\rangle \cdot
\nonumber\\[2mm]&&\hspace{-2mm}\cdot \langle\vec{P},\sqrt{s}|j_{\mu}^{c}(0)|\vec{P}',\sqrt{s'},1,0,1,m'\rangle
\langle\vec{P'},\sqrt{s'},1,0,1,m'|\vec{P_{\rho}},1,m_{\rho}\rangle\;,
\end{eqnarray}
where
$\langle\vec{P_{\pi}}|\vec{P},\sqrt{s}\rangle$ and $\langle\vec{P'},\sqrt{s'},1,0,1,m'|\vec{P_{\rho}},1,m_{\rho}\rangle$
 are the wave functions in the instant form of RQM,
\begin{eqnarray}\label{vf}
&&\hspace{-2mm} \langle\vec{P_{\pi}}|\vec{P},\sqrt{s}\rangle=
N_{c}\delta(\vec{P}-\vec{P_{\pi}})\varphi(s)\;,
\nonumber\\[2mm]&&\hspace{-2mm}
\langle\vec{P}',\sqrt{s'},1,0,1,m'|\vec{P_{\rho}},1,m_{\rho}\rangle=
N'_{c}\delta(\vec{P'}-\vec{P_{\rho}})\varphi^{1}_{1}(s')\delta_{m'm_{\rho}}\;,
\end{eqnarray}
\begin{equation}\label{1}
\varphi(s)= \sqrt[4]{s}k\psi({k}),  \quad \varphi^{1}_{1}(s')=
\sqrt[4]{s}k'\psi^{1}_{1}({k'})\;,
\end{equation}
and $\psi({k})$ and  $\psi^{1}_{1}({k'})$ are the wave functions satisfying the normalization condition

\begin{equation}\label{q1}
\int\psi^2(k)k^{2}dk=1\;,\quad
\int\left[\psi^{1}_{1}(k')\right]^2k'^{2}dk'=1 \;.
\end{equation}

Integrating over $\vec{P}$ and $\vec{P}'$ using delta functions, we obtain
\begin{eqnarray}\label{mat}
&&\hspace{-2mm}\langle
\vec{P_{\pi}}|j_{\mu}^{c}(0)|\vec{P_{\rho}},1,m_{\rho}\rangle =
\int\,d\sqrt{s}  d\sqrt{s'}\,\frac{N_c\,N'_c}{N\,N'}\, \varphi(s)
\varphi^{1}_{1}(s')\cdot
\nonumber\\[2mm]&&\hspace{-2mm}
\cdot
\langle\vec{P_{\pi}},\sqrt{s}|j^{c}_{\mu}(0)|\vec{P_{\rho}},\sqrt{s'},1,0,1,m_{\rho}\rangle\;.
\end{eqnarray}

The bra and ket vectors of the matrix element in the right-hand side of equality (\ref{mat}) physically describe a
system of two free particles and are transformed under a representation whose generators are independent of
the interaction. The current operator describes the transition between two-quark systems with interaction
and is transformed under another representation whose generators in RQM are interaction dependent.
Therefore, we cannot apply the parametrization procedure of Sec. 2. But we can treat the right-hand side of equality (\ref{mat}) as a functional
determined on the finite function space $\phi(s,s')=\varphi(s)\,\varphi^{1}_{1}(s')$ and interpret the matrix element of the
electromagnetic current as a distribution, i.e., as an object that makes sense only when it is in the integrand.

The transformation properties of the left- and right-hand sides of (\ref{mat})
coincide. Therefore, we can represent the nondiagonal parametrization of the matrix element in the
integrand in a form analogous to (\ref{A11A}),
(\ref{A111A}):
\begin{eqnarray}\label{A11A01}
&&\hspace{-2mm}
\frac{N_c\,N'_c}{N\,N'}\,\langle\vec{P_{\pi}},\sqrt{s}|{j^{c}_0}(0)|\vec{P_{\rho}},\sqrt{s'},1,0,1,m_{\rho}\rangle=\nonumber\\
[2mm]&& \hspace{-2mm}= \sum_{ \tilde{m_{\rho}}, l', k'}
D^{1}_{m_{\rho},\tilde{m_{\rho}}}(P_{\rho},w) \langle 1
\tilde{m_{\rho}} l'k' |0 0\rangle  Y_{l' k'}(\vec{q})
H^{0l'}_{01}(s,Q^2,s')\;,
\end{eqnarray}
\begin{eqnarray}\label{A111A02}
&&\hspace{-2mm}
\frac{N_c\,N'_c}{N\,N'}\,\langle\vec{P_{\pi}},\sqrt{s}|{j^c}^{1}_t(0)|\vec{P_{\rho}},\sqrt{s'},1,0,1,m_{\rho}\rangle = \nonumber\\
[2mm]&& \hspace{-2mm}= \sum_{\tilde{m_{\rho}}, t, k,n}
D^{1}_{m_{\rho}\tilde{m_{\rho}}}(P_{\rho},w)
 \langle 1\tilde{m_{\rho}} j n |0
0\rangle  \langle 1 t l k |j n\rangle  Y_{l
k}(\vec{q}) H^{1,l,j}_{01}(s,Q^2,s')\;,
\end{eqnarray}
where $H^{0l'}_{01}(s,Q^2,s'), H^{1,l,1}_{01}(s,Q^2,s')$ --
are form factors.

Substituting representations (\ref{A11A}), (\ref{A111A}) and (\ref{A11A01}), (\ref{A111A02}) in (\ref{mat}), we obtain
\begin{equation}\label{1a}
G^{{01}}_{01}(Q^2) = \int d\sqrt{s} d\sqrt{s'}
H^{01}_{01}(s,Q^2,s')\varphi(s)\varphi^{1}_{1}(s'),
\end{equation}

\begin{equation}\label{1b}
G^{{1,l,1}}_{01}(Q^2) = \int d\sqrt{s} d\sqrt{s'}
H^{1,l,1}_{01}(s,Q^2,s')\varphi(s)\varphi^{1}_{1}(s'),
\end{equation}

We note that the explicit form of the invariant functions $H^{01}_{01}(s,Q^2,s'), H^{1,l,1}_{01}(s,Q^2,s')$ is unknown in
general. We determine their explicit form using the MIA in \cite{KrT02}. This MIA consists in replacing the functions
$H^{01}_{01}(s,Q^2,s'),
H^{1,l,1}_{01}(s,Q^2,s')$ with free two-particle form factors $G^{01}_{01}(s,Q^2,s')$ and $G^{1,l,1}_{01}(s,Q^2,s')$.

\begin{equation}\label{Fc}
G^{01}_{01}(Q^2)=\int\,d\sqrt{s}
d\sqrt{s'}G^{01}_{01}(s,Q^2,s')\varphi(s)\varphi^{1}_{1}(s')\;,
\end{equation}
\begin{equation}\label{Fc1}
G^{1,l,1}_{01}(Q^2)=\int\,d\sqrt{s} d\sqrt{s'}
G^{1,l,1}_{01}(s,Q^2,s')\varphi(s)\varphi^{1}_{1}(s')\;.
\end{equation}

Therefore, taking (\ref{transition}) into account, we obtain the final expression for the experimentally measurable
transition form factor $F_{\pi\rho}(Q^2)$ in the MIA framework:

\begin{equation}\label{transition1}
F_{\pi\rho}(Q^2)=\sqrt{\frac{2}{3}} \,
\frac{1}{q({\tilde P}_{\pi}^{0}+{\tilde P}_{\rho}^{0})}\int\,d\sqrt{s}d\sqrt{s'}G_{01}^{111}(s,Q^2,s')\varphi(s)\varphi^{1}_{1}(s')\;.
\end{equation}

\section{Numerical calculation of the form factor $ F_{ \pi \rho}(Q^2)$}

When calculating the transition form factor by formulas \ref{transition1}) it is customary to use
the harmonic oscillator ground state function (see, e.g., \cite{CaG95,YuB07})),

\begin{equation}\label{volna1}
\psi(k)=\frac{2}{\pi^{1/4}\beta^{3/2}}\exp(-\frac{k^2}{2\beta^2})\;,
\end{equation}
where $\beta$  is a parameter.

We fix this parameter based on data for mean square
meson radii: $\beta = 0.278$ GeV for the pion \cite{HaN10} and $\beta =0.231$ GeV for the $\rho$-meson\cite{KrT03}.

To compare our results with those of other approaches, we choose the Sax form factors in the form

\begin{equation}\label{mmm}
G_{E}(Q^2)=e_{q}f_{q}(Q^2),\quad
G_{M}(Q^2)=(e_{q}+\kappa_{q})f_{q}(Q^2)\;,\quad
f_{q}(Q^2)=\frac{1}{1+\langle r^2_{q}\rangle Q^2/6)}\;,
\end{equation}
where $e_{q}$ is the quark charge, $\kappa_{q}$ is the anomalous magnetic moment of a quark expressed in the natural
units, and $\langle r^2_{q}\rangle$ is the mean square radius of a quark.

Let us briefly discuss the choice of parameter values for our numerical calculations. We use the value
$M=0.22\,$ GeV for masses of constituent $u$- and $\bar {d}$ quarks. This value is commonly used in modern relativistic
calculations (see, e.g., \cite{CaG95,YuB07}). We set the value of the mean square radius of a constituent quark to be
$\langle r^2_{q}\rangle=0.3/M^2$. This value was obtained in the model with spontaneous chiral symmetry breaking \cite{CaG95}, \cite{V90}.
We set the sum of anomalous magnetic moments of constituent quarks in the formula for the transition
form factor to be $\kappa_u + \kappa_{\bar d}= 0.09$  in accordance with the Gerasimov sum rules \cite{Ger95}.

\begin{figure}[h!]
\centering
\includegraphics[width=10.0cm]{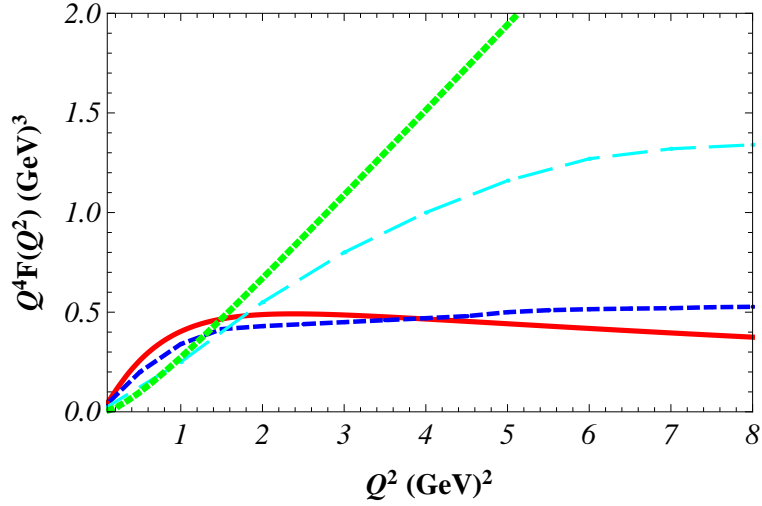}
\caption{The transition form factor of the process $\rho\rightarrow\pi\gamma^*$
calculated in various approaches: the
solid line is the result of calculating by formula (\ref{transition1}), short-dashed line is the result of calculating in
the framework of the light-front dynamics in \cite{CaG95}, long-dashed line is the result of calculating in the
framework of the light-front dynamics in \cite{YuB07}, and the dotted line is the result of calculating in the
vector meson dominance model (see, e.g., \cite{YuB07}).
\label{Plot:1a}}
\end{figure}

\section{Conclusion}
In the framework of instant form of relativistic quantum mechanics
the calculation of the transition form factor for the radiative
decay $\rho\rightarrow\pi\gamma^*$ is performed. For the
calculation the original method of the parametrization of the
matrix element of the transition current is used. In the modified
relativistic impulse approximation we obtained analytic expression
for the transition form factor. The transition form factor is
written via the two-quark wave functions and the so-called free
two-particle form factor which enters in the matrix element of the
transition current between two systems of free quarks with the
quantum numbers of pion and $\rho$-meson. In contrast to the
standard impulse approximation, the modified relativistic impulse
approximation does not break the Lorentz covariance and the
current conservation law. The numerical calculation results in the
framework of our approach are close to the results obtained in the
light-front dynamics and agree with the vector meson dominance
model.

}


\begin{thebibliography}{99}

\bibitem {Dir49}
P. A. M. Dirac, \emph{Rev. Modern Phys} {\bf 21}, 392--399 (1949).

\bibitem {LeS78}
H. Leutwyler and J. Stern, \emph{Ann. Phys} {\bf 112},94--164 (1978).

\bibitem {KeP91}
B. D. Keister and W. N. Polyzou, \emph{Advances in Nuclear Physics} {\bf 20}, 225--479 (1991).

\bibitem {Coe92}
F. Coester, \emph{ Progr. Part. Nucl. Phys} {\bf 29}, 1--32 (1992).

\bibitem {GiG02}
E. Gilman and F. Gross, \emph{J. Phys. G} {\bf 28}, 37--116 (2002); arXiv:nucl-th/0111015v1 (2001).

\bibitem {KrT09}
A. F. Krutov and V. E. Troitsky, \emph{Phys. Part. Nucl.} {\bf 40}, 136--161 (2009).

\bibitem {PoE11}
W. N. Polyzou, Ch. Elster, W. Gl¨ockle, J. Golak, Y. Huang, H. Kamada, R. Skibi´nski, and H. Witala, \emph{Few-Body
Systems} {\bf 49},129--147 (2011).

\bibitem {Pol89}
W. N. Polyzou, \emph{Ann. Phys} {\bf 193}, 367--418 (1989).

\bibitem {KrT05}
A. F. Krutov and V. E. Troitsky, \emph{Theor. Math. Phys.} {\bf 143}, 704--719 (2005).

\bibitem {TrS69}
V. E. Troitsky and Yu. M. Shirokov, \emph{Theor. Math. Phys.} {\bf 2}, 164--170 (1969).

\bibitem {KrT02}
A. F. Krutov and V. E. Troitsky, \emph{Phys. Rev. C} {\bf 65}, 045501 (2002); arXiv:hep-ph/0101327v1 (2001).

\bibitem {BaK95}
E. V. Balandina, A. F. Krutov, and V. E. Troitsky, \emph{Theor. Math. Phys.} {\bf 103},381--389  (1995).

\bibitem {BaK96}
E. V. Balandina, A. F. Krutov, and V. E. Troitsky, \emph{J. Phys. G} {\bf 22}, 1585--1592 (1996).

\bibitem {KrT99}
A. F. Krutov and V. E. Troitsky, \emph{JHEP} {\bf 9910}, 028 (1999).


\bibitem {KrT01}
A. F. Krutov and V. E. Troitsky, \emph{Eur. Phys. J. C} {\bf 20}, 71--76  (2001).


\bibitem {KrT98}
A. F. Krutov and V. E. Troitsky, \emph{Theor. Math. Phys.} {\bf 116}, 907--913 (1998).

\bibitem {KrN13}
 A. F. Krutov, M. A. Nefedov, and V. E. Troitsky, \emph{Theor. Math. Phys.} {\bf 174:3},331--342 (2013).


\bibitem {KrTT09}
A. F. Krutov, V. E. Troitsky, and N. A. Tsirova, \emph{Phys. Rev. C} {\bf 80},  055210  (2009); arXiv:0910.3604v2 [nucl-th]
(2009).

\bibitem {TrT13}
S. V. Troitsky and V. E. Troitsky, \emph{ Phys. Rev. D} {\bf 88}, 093005  (2013); {\bf 91}, 033008 (2015); arXiv:1501.02712v2
[hep-ph] (2015).


\bibitem {ChS63}
A. A. Cheshkov and Yu. M. Shirokov, \emph{Soviet Phys. JETP} {\bf 17}, 1333--1339 (1963).


\bibitem{KrP15} A. F. Krutov, R. G. Polezhaev, and V. E. Troitsky, \emph{ Theor. Math. Phys.} {\bf 184:2}, 1148--1162 (2015).


\bibitem {CaG95}
F. Cardarelli, I. L. Grach, I. M. Narodetskii, G. Salm´e, and S. Simula, \emph{Phys. Lett. B} {\bf 359}, 1--7 (1995); arXiv:
nucl-th/9509004v2 (1995).

\bibitem {YuB07}
J. Yu, B.-W. Xiao, and B.-Q. Ma, \emph{J. Phys. G} {\bf 34}, 1845--1860 (2007); arXiv:0706.2018v1 [hep-ph] (2007).

\bibitem{HaN10} H.E.~Haber, K.~Nakamura, \emph{J. Phys. G. Nucl. Part. Phys.} {37}, 075021 (2010).

\bibitem {KrT03}
A. F. Krutov and V. E. Troitsky, \emph{ Phys. Rev. C} {\bf 68}, 018501 (2003); arXiv:hep-ph/0210046v1 (2002).


\bibitem {V90}
U. Vogl, M. Lutz, S. Klimt, and W. Weise, \emph{Nucl. Phys. A} {\bf 516:3},469--495 (1990); B. Povh and J. H¨ufner, \emph{Phys. Lett. B} {\bf 245:3--4},653--657 (1990);S. M. Troshin and N. E. Tyurin, \emph{Phys. Rev. D} {\bf 49:9},4427--4433 (1994).


\bibitem {Ger95}
S. B. Gerasimov, \emph{Phys. Lett. B} {\bf 357}, 666--670 (1995).

\end{thebibliography}
\end{document}